\pdfoutput=1
\documentclass[iop,apj]{emulateapj}

\usepackage{amsmath}
\usepackage{bm}
\usepackage{natbib}
\usepackage{url}
\usepackage{graphicx}
\usepackage[breaklinks,colorlinks,
  urlcolor=blue,citecolor=blue,linkcolor=blue]{hyperref}
\newcommand{\planck}{{\sl Planck}}

\newcommand{\vev}[1]{\langle#1\rangle}

\shorttitle{Imaging parity-violation in the CMB}
\shortauthors{Contaldi}

\begin{document}

\preprint{Imperial/TP/2015/CC/4} \vskip 0.2in

\title{Imaging
  parity-violating modes in the CMB}

\author{Carlo~R.~Contaldi}
\affil{Theoretical Physics Group, Blackett Laboratory,
  Imperial College London, South Kensington Campus, London, SW7 2AZ,
  UK}
  \email{c.contaldi@imperial.ac.uk}     

\begin{abstract}
  Correlations of polarization components in the coordinate frame are
  a natural basis for searches of parity-violating modes in the Cosmic
  Microwave Background (CMB). This fact can be exploited to build
  estimators of parity-violating modes that are {\sl local} and robust
  with respect to partial-sky coverage or inhomogeneous weighting. As
  an example application of a method based on these ideas we develop a
  peak stacking tool that isolates the signature of parity-violating
  modes. We apply the tool to {\sl Planck} maps and obtain a
  constraint on the monopole of the polarization rotation angle
  $\alpha < 0.72$ degrees at $95\%$ We also demonstrate how the tool
  can be used as a local method for reconstructing maps of direction
  dependent rotation $\alpha(\hat {\bm n})$.
\end{abstract}


\section{Introduction}

The CMB provides a powerful test of deviations from standard
physics. The energy fluctuations and polarization of photons released
at last scattering carry information that is directly linked via
simple, linear evolution to the primordial state of super horizon
perturbations of the metric. In addition,  CMB photons have traveled
over cosmological distances to reach us, probing the nature
of space-time along the way. The polarization of the CMB in, 
particular, is sensitive to a number of  parity-violating mechanisms due to the opposite
parity of $E$ and $B$-modes
\citep{Kamionkowski:1996zd,Seljak:1996gy,Kamionkowski:1996ks}.


These mechanisms can act either during the generation of primordial
perturbations or during the propagation of the CMB photons through
space at later times. The first possibility includes models of chiral
gravity \citep{Lue:1998mq,2005PhRvD..71f3526A,Contaldi:2008yz,Sorbo:2011rz}
that explicitly break the parity between left and right handed
primordial gravitational waves imprinted in super-horizon scales
during inflation. The second includes the presence of parity-violating
primordial magnetic
fields \citep{1996ApJ...469....1K,PhysRevLett.87.101301} or mechanisms
that induce cosmic
birefringence \citep{1998PhRvL..81.3067C,Lue:1998mq,Feng:2004mq,Li:2008tma} through the
presence of new fields and interactions. Another possibility is the
presence of spatial anisotropies during evolution of
perturbations \citep{Bartolo:2015dga}.

The simple consequence of all these mechanisms is that they result in
correlations between the $E$ and $B$-modes of the CMB polarization and
between the $B$-mode and total intensity $T$. These correlations
cannot exist unless parity is violated. The hope is therefore that a significant
measurement of $EB$ or $TB$ correlations would be a
``clean'' indication of non-standard physics. The difficulty however
is that these signals can also arise due to any experimental
systematic that coherently rotates the polarization signal with
respect to the true polarization of CMB photons. This can happen in
experiments via either optical effects or mis-calibration of the
polarization direction or efficiency of detectors.

With the advent of high signal-to-noise measurements of CMB
polarization \citep{Ade:2014afa,Ade:2014xna,Ade:2015tva,Naess:2014wtr,Keisler:2015hfa}
we have entered an era where extensive test of parity-violating
effects can be carried out along with a requirement for increasingly
precise tests of polarization affecting systematics. Indeed, much work
has gone into defining estimators of parity-violating statistics and
their interaction with the experimental polarization calibration
procedures \citep{Cabella:2007br,2009PhRvD..80d3522P,2009ApJ...705..978B,Komatsu:2010fb,Gruppuso:2015xza,Ade:2015cao}. 

Estimation of the parity-violating signal has focused on the
definition of optimal estimators based on the spherical harmonic
expansion of the CMB Stokes parameters $I$, $Q$, and $U$ into $T$,
$E$, and $B$-mode spherical harmonic coefficients. The focus on 
working in the $T$, $E$, $B$ basis is a naturally justified one since the
correlation of these modes is where the signal of parity-violation is
explicit. In contrast, it is not immediately obvious that correlations
in the original $I$, $Q$, and $U$, coordinate frame modes can uniquely distinguish
parity-violating signals as there is no one-to-one mapping of
$Q$ and $U$ modes to the parity-sensitive $E$ and $B$ modes. Working in harmonic space
however, involves well-known complications due to the fact that the
transformations involved are non-local. When a partial-sky coverage is
imposed by scanning limitations or contamination by
foregrounds the orthonormality of basis functions is reduced. This leads
to mixing of power at different angular scales {\sl and} mixing of $E$
and $B$-modes. To deal with these unwanted effects one has to resort
to complicated methods to {\sl statistically} isolate the original
harmonic modes \citep{Hivon:2001jp,Bunn:2002df,Smith:2006vq}.

In this work we point out that simple combinations
of cross-correlations of Stokes parameters in the original coordinate
frame are parity-violation sensitive and we show how these can be
applied to data to yield an intuitive compression of the signals for
parity-violating tests. Estimators based on coordinate frame
correlations are simpler, and, in principle, more robust with respect
to inhomogeneous noise, partial-sky coverage effects, and any polarization
systematics. In fact, aside from efforts to detect non-standard
physics, they can provide versatile tests of experimental systematics.

This {\sl paper} is organised as follows. In section~\ref{sec:functions} we
review how angular correlation functions of the Stokes parameters are
related to parity-even and parity-odd angular power spectra. In
section~\ref{sec:stacks} we move to the small-angle limit and show how
peak stacking can be used to obtain compressions of the signal that
isolate parity-violating correlations. We validate our peak stacking
tools using simulations of the CMB sky that contain varying degrees of
parity-violation in the form of an overall polarization rotation. In
section~\ref{sec:appl} we use the tool to search for any
parity-violating signal in the {\sl Planck} maps. We discuss our
results in section~\ref{sec:disc}. 

\section{Polarization Correlation Functions}\label{sec:functions}

The polarization of CMB photons on the full sky is described in terms
of Stokes parameters, $T$, $Q$, and $U$. The scalar, total intensity
component $T$ is expanded in spherical harmonics as
\begin{equation}
  T(\hat{\bm n}) = \sum_{\ell m}a_{\ell m}^{}Y_{\ell m}(\hat{\bm n})\,,
\end{equation}
where $\hat{\bm n}$ is the unit direction vector on the sky. The Stokes
parameters determining the direction of polarization on the sky are
expanded on the basis of spin-2 spherical harmonics
\citep{Zaldarriaga:1996xe}
\begin{equation}
  (Q\pm iU)(\hat{\bm n}) = \sum_{\ell m}(a_{\ell m}^E\mp i\,a_{\ell
    m}^B )\, _{\mp 2}Y_{\ell m}(\hat{\bm n})\,.
\end{equation}
The $E$ and $B$-modes transform distinctly under parity
transformations $(Q\pm iU)(\hat{\bm n})\to (Q\mp iU)(-\hat{\bm
  n})$ with the gradient-like mode $a_{\ell m}^E$ transforming with
parity $(-1)^\ell$ and curl-like mode $a_{\ell m}^B$ transforming as
$(-1)^{\ell +1}$. Given statistical isotropy the correlations in the
expanded modes are diagonal and can be split into correlations that are
insensitive to parity transformations
\begin{align}
    \vev{a_{\ell m}^Ta_{\ell'
        m'}^{T\star}}&=\delta_{\ell\ell'}\delta_{mm'}C_\ell^{T}\,,\label{eq:cl1}\\
     \vev{a_{\ell m}^Ta_{\ell'
         m'}^{E\star}}&=\delta_{\ell\ell'}\delta_{mm'}C_\ell^{TE}\,,\label{eq:cl2}\\
     \vev{a_{\ell m}^Ea_{\ell'
         m'}^{E\star}}&=\delta_{\ell\ell'}\delta_{mm'}C_\ell^{E}\,,\label{eq:cl3}\\
         \vev{a_{\ell m}^Ba_{\ell'
         m'}^{B\star}}&=\delta_{\ell\ell'}\delta_{mm'}C_\ell^{B}\,,\label{eq:cl4}
\end{align}
and ones that only exist if the are parity-violating modes in the
pattern
\begin{align}
  \vev{a_{\ell m}^Ta_{\ell'
      m'}^{B\star}}&=\delta_{\ell\ell'}\delta_{mm'}C_\ell^{TB}\,,\label{eq:cl5}\\
    \vev{a_{\ell m}^Ea_{\ell'
         m'}^{B\star}}&=\delta_{\ell\ell'}\delta_{mm'}C_\ell^{EB}\,.\label{eq:cl6}
\end{align}

The complex spin-2 field $P=Q+iU$ is a convenient measure for defining
correlations in coordinate space. When considering correlations of the
polarization between two lines of sight $\hat{\bm n}_1$ and
$\hat{\bm n}_2$ we have to take into account that the polarization
is defined with respect to a local basis, typically based on the local
meridian. Rotating the polarization, in both directions, to a common
frame defined by the geodesic connecting the two points on the sky
subtending angles $\gamma_1$ and $\gamma_2$ with the original $x$-axis
at both points
\begin{equation}\label{eq:geodesic}
  \bar P(\hat{\bm n}_j) = e^{2i\gamma_j}\, P(\hat{\bm
    n}_j)\,,
\end{equation}
with $j=1,2$, one obtains a polarization definition that simplifies the relation 
between coordinate and harmonic space based correlations
\citep{Ng:1997ez,Chon:2003gx}
\begin{align}
  \vev{T(\hat{\bm n}_1)T(\hat{\bm n}_2)} &= \sum_\ell
    \frac{2\ell+1}{4\pi} \,C_\ell^{T}\,P_\ell(\beta)\,\label{eq:corr1}\\
    \vev{T(\hat{\bm n}_1)\bar P(\hat{\bm n}_2)}&= \sum_\ell
      \frac{2\ell+1}{4\pi} \,\left(C_\ell^{TE}-iC_\ell^{TB}\right)\,d^\ell_{20}(\beta)\,\label{eq:corr2}\\
      \vev{P(\hat{\bm n}_1)\bar P(\hat{\bm n}_2)}&= \sum_\ell
    \frac{2\ell+1}{4\pi}
    \,\left(C_\ell^{E}-C_\ell^{B}-2iC_\ell^{EB}\right)\nonumber\\
    &\times d^\ell_{2-2}(\beta)\,\label{eq:corr3}\\
  \vev{\bar P^\star(\hat{\bm n}_1)\bar P(\hat{\bm n}_2)}&=\sum_\ell
    \frac{2\ell+1}{4\pi} \,\left(C_\ell^{E}+C_\ell^{B}\right)\,d^\ell_{22}(\beta)\,\label{eq:corr4}
\end{align}
where $\beta$ is the cosine of the angle between the two directions and
$d^\ell_{mm'}$ are the reduced Wigner rotation matrices
\citep{varshalovich1988quantum}. These relations can be inverted to yield angular
power spectra as functions of coordinate space correlations by using
the orthogonality of the basis functions in the full-sky limit. Note
that the $C_\ell^E$ and $C_\ell^B$ spectra contribute to the real part
of both equations (\ref{eq:corr3}) and (\ref{eq:corr4}) but are
multiplied by different basis functions. This is the reason why
cut-sky effects have to be dealt with even when considering coordinate
space correlation functions - the $E$ and $B$ mode contribution cannot
be disentangled. The parity-violating components $TB$ and $EB$ however,
only contribute to the imaginary parts of the correlations 
(\ref{eq:corr2}) and (\ref{eq:corr3}). In practice this means that we
can use correlations obtained in the coordinate frame to isolate the
signal due to parity-violating modes. Any estimate that does not rely
on a harmonic transform greatly simplifies the problem of partial-sky
coverage and inhomogeneous weighting.

The disadvantage is that the computation of correlation functions
directly in the coordinate frame requires ${\cal O}(N_p^2)$
operations, where $N_p$ is the number of pixels in the map. Typically
even methods that make use of the coordinate frame correlations 
(\ref{eq:corr1})-(\ref{eq:corr4}) to simplify the effect of masking
and noise weighting \citep{Chon:2003gx} actually estimate the
correlations by transforming to the harmonic space. This is done in
order to make use of the fast Fourier transforms in longitude.

In the following we will show how considering a subset of all possible
correlations, those constrained to the neighbourhood of peaks in the
Stokes parameters, can make use of the equations
(\ref{eq:corr1})-(\ref{eq:corr4}) to isolate the coordinate frame
signal of parity-violating modes. We will leave the development of a
more general framework exploiting the unconstrained correlations in
the full-sky for future work.

\section{Peak stacks}\label{sec:stacks}

Peak stacking, or the averaging of a map in the neighbourhood of peaks
in the map, has been used to determine the robustness of
polarization measurements since large scale, polarization sensitive
surveys appeared \citep{Komatsu:2010fb}. Recently they have been used to
analyse more detailed morphology of the CMB around acoustic peaks in an
attempt to constrain deviations from statistical isotropy
\citep{Ade:2015hxq}. Peak stacking provides a powerful visualization of
the constrained correlation pattern around an acoustic peak of the
CMB. It constitutes an intuitive tool that highlights the nature of the
polarization as being precisely that expected from acoustic
oscillations at last scattering. In the following we consider stacks
around peaks in each of the Stokes parameters of polarised maps and
show how these can be used to obtain parity-sensitive stacks that have
simple relations to the parity-violating spectra $C_\ell^{TB}$ and
$C_\ell^{EB}$. 

We will follow the notation and conventions used in
\cite{Komatsu:2010fb} and consider explicit contributions $Q$ and $U$
in $P$ and expand the phase factor in equation~(\ref{eq:geodesic}) in
terms of sines and cosines. We develop the framework in the
small-angle limit by considering the tangential projection, in
coordinates ($x$, $y$), around a point in the sky along direction
$\hat{\bm n}$. In this case the spin-$0$, and $\pm 2$ spherical
harmonics are replaced by two dimensional plane waves and their
derivatives
\begin{align}
  Y_{\ell m}&\to
  e^{i\bm{\ell}\cdot\bm{\theta}}\,\\
  _\pm Y_{\ell m}&\to 
  -e^{\mp 2i(\phi-\varphi)}\,e^{i\bm{\ell}\cdot\bm{\theta}}\,,
\end{align}
where ${\bm \ell} = $ ($\ell_1$, $\ell_2$) is the wave vector in the
plane-wave expansion with $\ell_1 = \ell \cos (\varphi) $ and $\ell_2 =
\ell\sin(\varphi)$ and $\bm \theta =$ ($\theta_1$, $\theta_2$) is the position
vector in the tangential projection with $\theta_1 = \theta \cos
(\phi) $ and $\theta_2 =
\theta\sin(\phi)$.

The total intensity field $T$ can now be expanded in the small-scale
limit in the space of two-dimensional plane-waves
\begin{equation}
 T(\bm\theta) = \int \frac{d^2\bm\ell}{(2\pi)^2}\,T_{\bm\ell}\,e^{i\bm{\ell}\cdot\bm{\theta}}\,.
\end{equation}
Following the conventions of \cite{Komatsu:2010fb,Kamionkowski:1996ks} we also expand the $Q$ and $U$
Stokes fields as
\begin{align}
  Q(\bm\theta) &= - \int \frac{d^2\bm\ell}{(2\pi)^2}\left[E_{\bm\ell}\cos(2\varphi)-B_{\bm\ell}\sin(2\varphi))\right]e^{i\bm{\ell}\cdot\bm{\theta}}\,,\\
  U(\bm\theta) &= - \int \frac{d^2\bm\ell}{(2\pi)^2}\left[E_{\bm\ell}\sin(2\varphi)+B_{\bm\ell}\cos(2\varphi))\right]e^{i\bm{\ell}\cdot\bm{\theta}}\,.
\end{align}
 The coefficients $E_{\bm \ell}$ and $B_{\bm \ell}$
are the gradient and curl-like modes of the polarization component.

When stacking, or averaging fields around peak
locations, we are calculating constrained (or biased) correlation
functions \citep{Bardeen:1985tr}. We can write the stacking of a field $X$ at
peaks in field $Y$ in the ensemble limit as \citep{Komatsu:2010fb}
\begin{equation}
  \vev{X}(\bm\theta) = \frac{1}{N^Y_{pk}}\int_M
  d\Omega\,\vev{n^Y_{pk}(\hat {\bm n})X(\hat {\bm n}+\bm\theta)}\,,
\end{equation}
where $N^Y_{pk}$ is the total number of peaks in an area of the sky
defined by the mask $M$ and $n^Y_{pk}(\hat {\bm n})$ is the number
density of peaks in the field $Y$ in the direction $\hat {\bm n}$.

Introducing the {\sl dimensionless} density contrast of peaks around the average peak
density $\delta^Y_{pk} = n^Y_{pk}/{\bar n}^Y_{pk} -1$, we can write the
stacking as
\begin{equation}
    \vev{X}(\bm\theta) = \frac{1}{4\pi\,f_{\rm sky}}\int_M
  d\Omega\,\vev{\delta^Y_{pk}(\hat {\bm n})X(\hat {\bm n}+\bm\theta)}\,,
\end{equation}
where $f_{\rm sky}$ is the fraction of the sky defined by the coverage
mask and $N^Y_{pk}= 4\pi\,f_{\rm sky}\,\bar n^Y_{pk}$.

The effect of peak biasing can be calculated from first principles if
$Y$ is a Gaussian random field \citep{Kaiser:1984sw,Desjacques:2008vf,Komatsu:2010fb}. The
density contrast can be written as a bias of the underlying field as
\begin{align}\label{eq:delta}
  \delta^Y_{pk}(\hat {\bm n}) = \left[
  b^Y_\nu-b^Y_\zeta(\partial^2_1+\partial^2_2)\right] Y (\hat {\bm n})\,,
\end{align}
where $b_\nu$ is a scale-independent bias due to the selection of
peaks of height $\nu$, in units of the field's standard deviation
$\sigma_0$ and $b_\zeta$ is a scale-dependent bias due to
the selection involving the curvature of the field. When considering
stack averages therefore, we are actually evaluating a quantity that is
second order in the statistics of the fields involved despite the
average being first order in field units. Thus, despite the field
having zero mean in the unconstrained limit, the constrained mean is
not expected to vanish. The averages are related to the
unconstrained correlation functions of the fields via a convolution
with filters determined by the biasing functions such as in equation~(\ref{eq:delta}).

Introducing the short-hand
$\vev{\delta_{\rm pk}^Y\,X}(\bm\theta)\equiv \vev{X^Y}$ and making use of the
definition of the angular cross-correlations relations between
statistically isotropic fields $X$ and $Y$
\begin{equation}
  \vev{X_{\bm\ell'}Y_{\bm\ell}} = (2\pi)^2 \delta^{(2)}(\bm\ell'-\bm\ell)\,C_\ell^{XY}\,,
\end{equation}
with $X$ and $Y$ belonging to the set $(T$, $Q$, $U)$, the constrained mean of the
temperature field about temperature peaks is related to the total
intensity angular power spectrum as\footnote{For a detailed summary of
how $\delta_{pk}$ is calculated we refer to Appendix B of \cite{Komatsu:2010fb}.}
\begin{equation}
\vev{T^T} =\int\frac{\ell d\ell}{2\pi} \,f_\ell^T \,C_\ell^{T}\, J_0(\ell\theta)\,,
\end{equation}
where above and in the following $J_n(x)$ are Bessel
functions of order $n$ arising from the integration of the expressions
over the plane wave azimuthal angle $\varphi$ and we have introduced
\begin{equation}
  f_\ell^X = b_\nu^X + b_\zeta^X\,\ell^2\,,
\end{equation}
as the Fourier domain counterpart of equation (\ref{eq:delta}).

Similarly, one can show
that suitably rotated combinations of $Q$ and $U$ constrained means
are related to the two cross-correlation angular power spectra
\begin{align}
  \vev{Q_r}&\equiv\vev{Q^T}\cos(2\phi) + \vev{U^T}\sin(2\phi)
  \nonumber\\
  &=\int\frac{\ell
     d\ell}{2\pi} \,f_\ell^T \,C_\ell^{TE}\, J_2(\ell\theta)\,,\label{eq:qr}\\
   \vev{U_r}&\equiv -\vev{Q^T}\sin(2\phi) + \vev{U^T}\cos(2\phi)
   \nonumber\\
   &=\int\frac{\ell
     d\ell}{2\pi} \,f_\ell^T \,C_\ell^{TB}\, J_2(\ell\theta)\,.\label{eq:ur}
\end{align}
The radial polarization projection of equation (\ref{eq:radial}) removes the
azimuthal dependence of the polarization pattern around the peaks and
gives a distinct signal from parity-violating modes arising from any
non-vanishing $C_\ell^{TB}$ contribution. It is therefore a useful
compression of the data along with a powerful visual check for the
presence of any anisotropic or polarization rotation systematics
\citep{Komatsu:2010fb,Ade:2015hxq}.

\begin{figure*}
\begin{center}\begin{tabular}{c}
    \includegraphics[width=12cm]{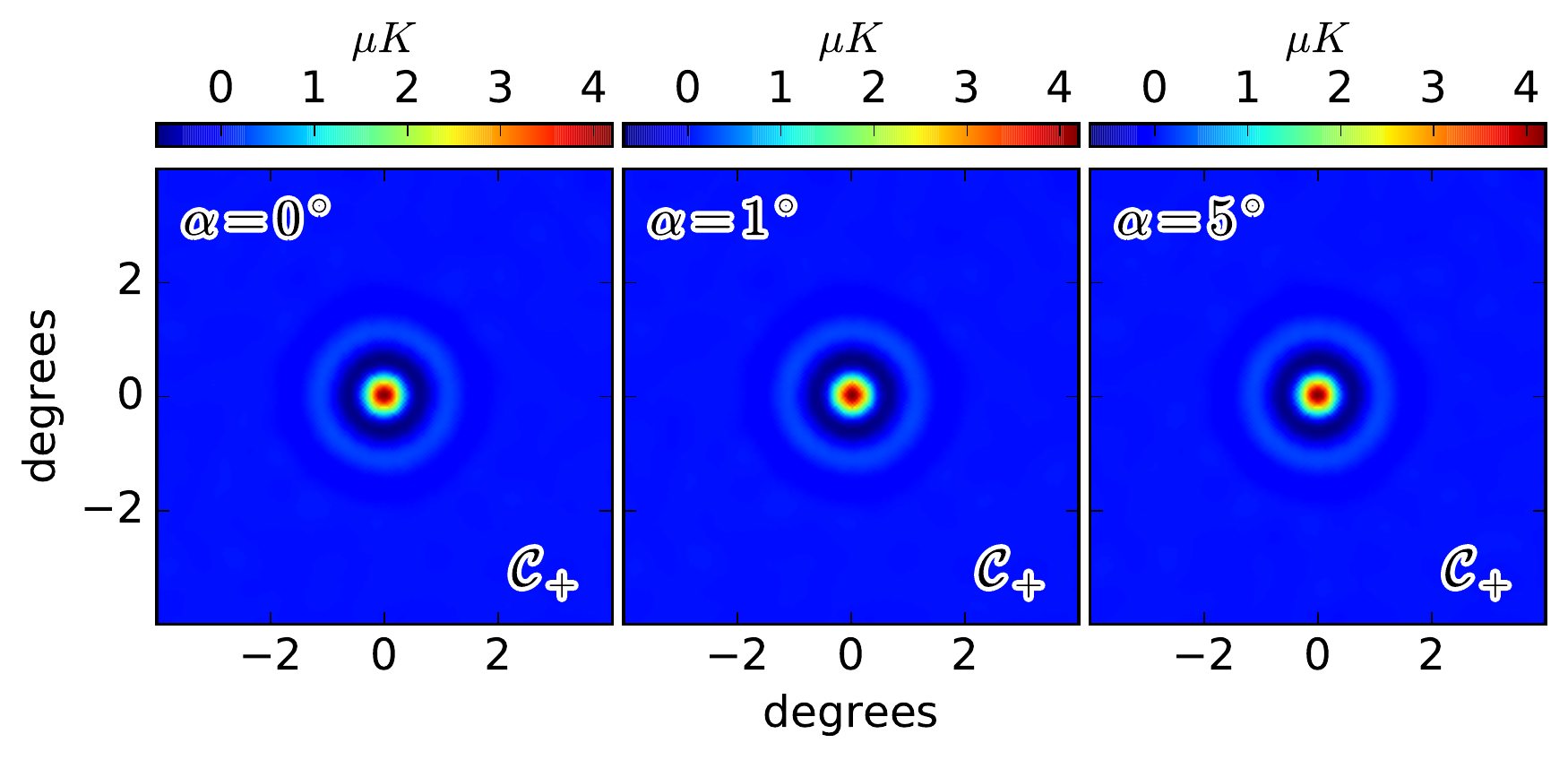}\\
    \includegraphics[width=12cm]{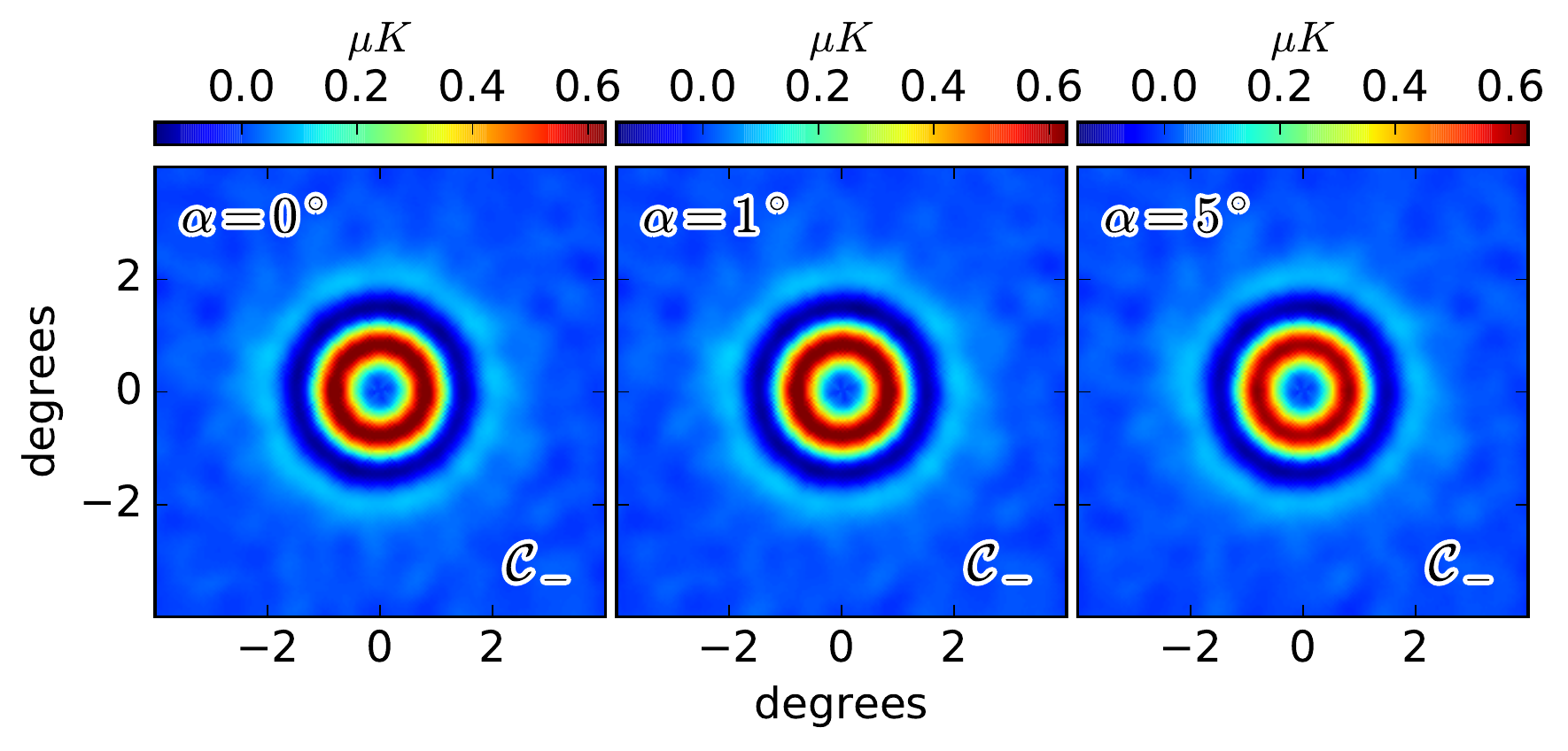} \\
    \includegraphics[width=12cm]{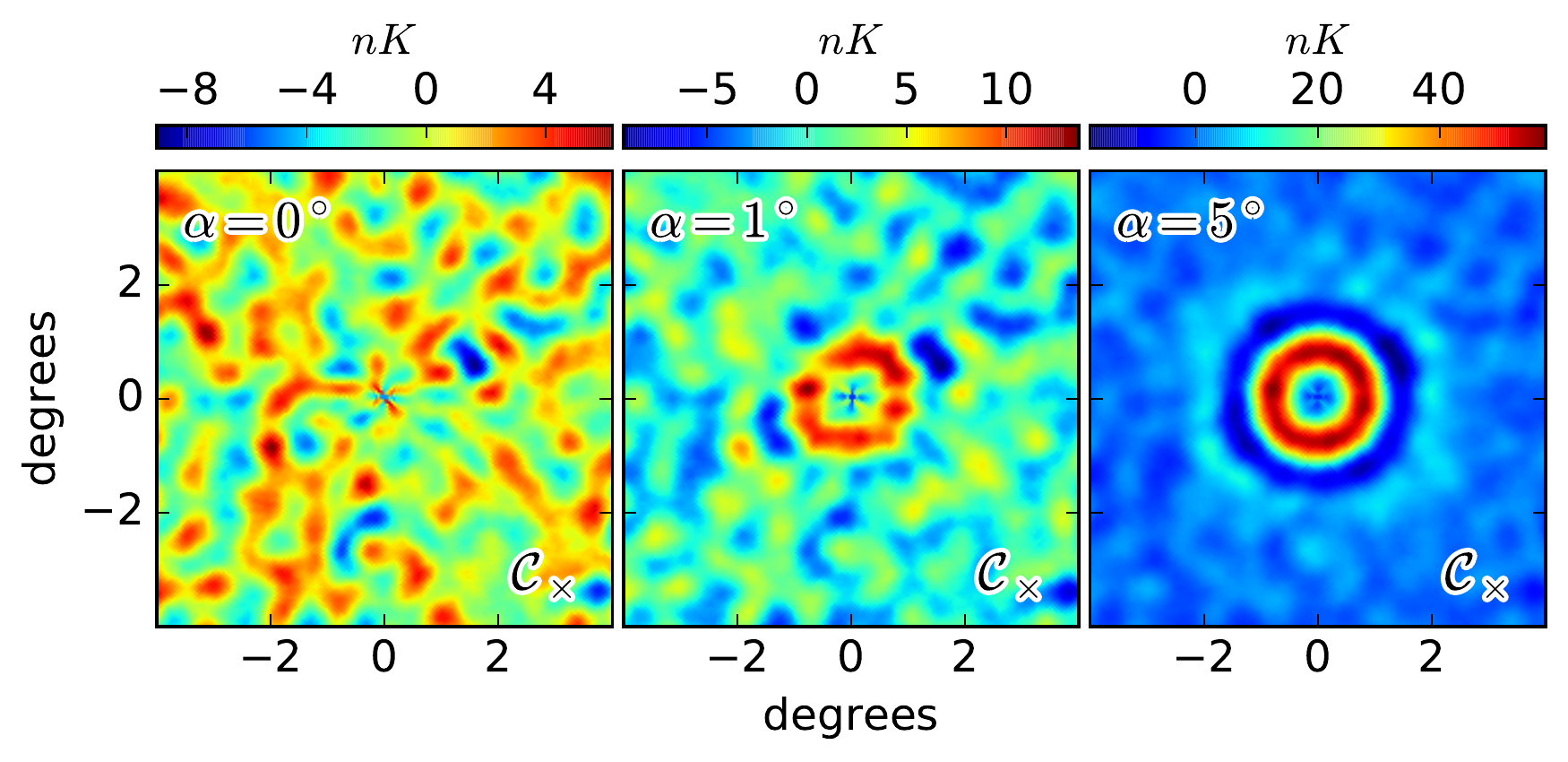}\\
    \includegraphics[width=12cm]{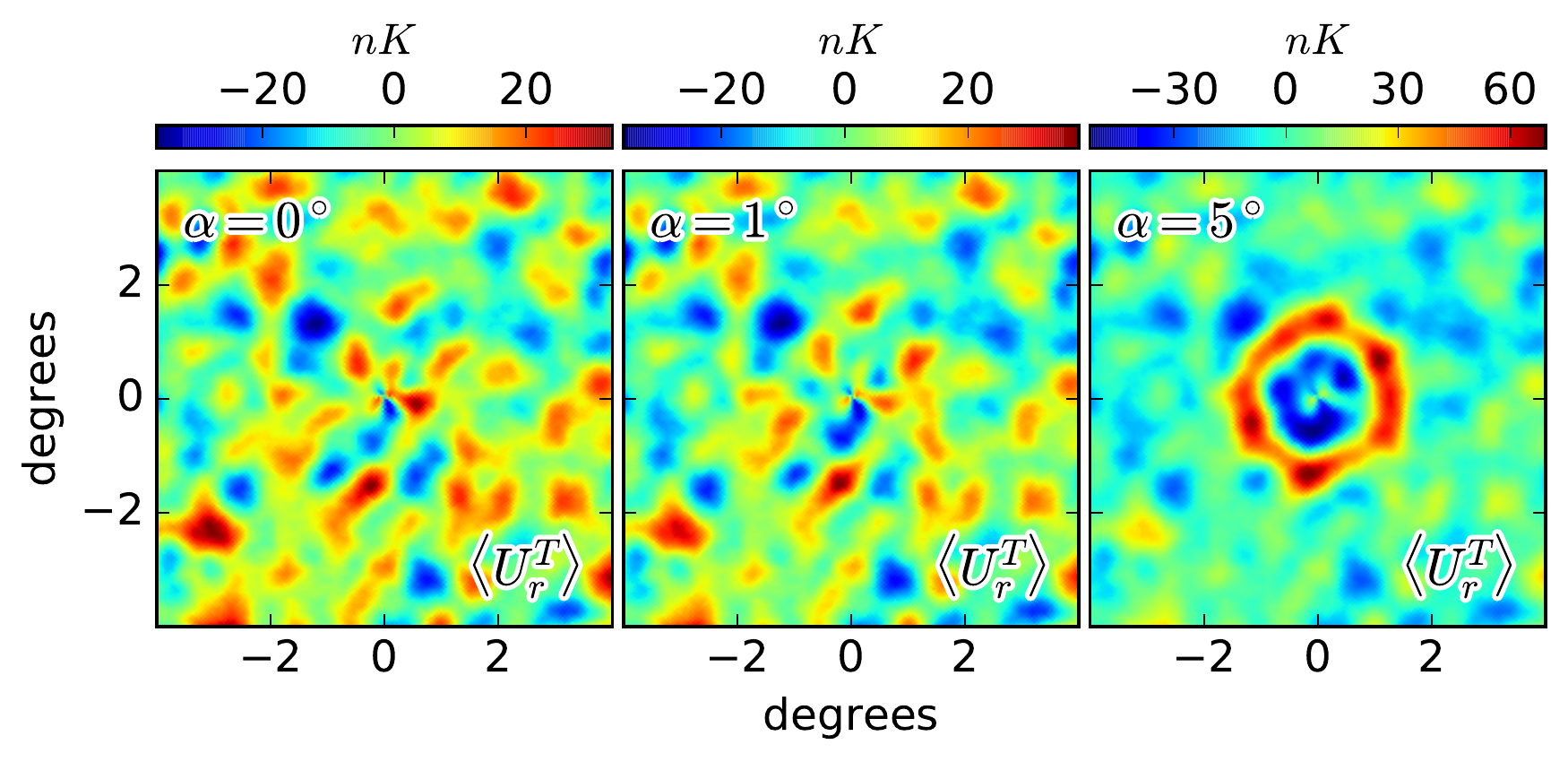}
\end{tabular}
\caption[]{Results from {\sl Planck}-like simulations smoothed to 30
  arcminute resolution with rotation angles $\alpha=0$,
  $1$, and $5$ degrees. From {\sl top} to {\sl bottom}; ${\cal C}_+$,
  ${\cal C}_-$, ${\cal C}_\times$, and $\vev{U^T_r}$ showing the emergence of the
  parity-violating signal in the two estimators sensitive to $EB$
   and $TB$  modes. The colour scales are
selected to enhance the contrast in each of the images.}
\label{fig:simulations}
\end{center}
\end{figure*}

The constrained mean around $Q$ and $U$ peaks can also yield useful
projections of the polarization data that distinguish between
parity-conserving and parity-violating modes. Following a similar 
derivation used for the expressions above, one can obtain  the following useful combinations of averages
around $Q$ and $U$ peaks
\begin{widetext}
\begin{align}
  {\cal C}_+(\bm\theta) &\equiv \vev{
    Q^Q+U^U}\nonumber\\
  &=\frac{1}{2}\int \frac{\ell
    d\ell}{2\pi} \,f_\ell^+ \,\left(C_\ell^{E}
  +C_\ell^{B}\right)\,J_0(\ell\theta)
  + f_\ell^- \,\left(C_\ell^{E}
  -C_\ell^{B}\right)\,J_4(\ell\theta)\,\cos(4\phi)
  -2\,f_\ell^-
  \,C_\ell^{TB}\,J_4(\ell\theta)\,\sin(4\phi)\,, \label{eq:ps1}\\
    {\cal C}_-(\bm\theta) &\equiv \vev{
  Q^U+U^Q}\sin(4\phi)+ \vev{
      Q^Q-U^U}\cos(4\phi)\nonumber\\
    &=\frac{1}{2}\int \frac{\ell
    d\ell}{2\pi} \,f_\ell^+ \,\left(C_\ell^{E}
  -C_\ell^{B}\right)\,J_4(\ell\theta)+ f_\ell^- \,\left(C_\ell^{E}
  +C_\ell^{B}\right)\,J_0(\ell\theta)\,\cos(4\phi)\,, \label{eq:ps2}\\
  {\cal C}_\times(\bm\theta) &\equiv \vev{
  Q^U+U^Q}\cos(4\phi)- \vev{
  Q^Q-U^U}\sin(4\phi)\nonumber\\
  &=\frac{1}{2}\int \frac{\ell
    d\ell}{2\pi} \,4\,f_\ell^+ \,C_\ell^{EB}\,J_4(\ell\theta)
  - f_\ell^- \,\left(C_\ell^{E}
  +C_\ell^{B}\right)\,J_0(\ell\theta)\,\sin(4\phi)\,, \label{eq:ps3}\\
    {\cal C}_0(\bm\theta) &\equiv \vev{
  Q^Q}-\vev{U^U} =0\,,\label{eq:ps4}
\end{align}
\end{widetext}
where we have introduced the notation $f_\ell^\pm\equiv f_\ell^Q\pm
f_\ell^U$.

Peak bias functions for the polarization fields are harder to
calculate {\sl a priori} since the original, unrotated $Q$ and $U$ are
generally correlated but in the limit of large $N_{pk}$ we expect to
recover  $f_\ell^Q=f_\ell^U\to f_\ell^P$ \footnote{Alternatives
would be to either stack around peaks in $|P|=\sqrt{Q^2+U^2}$ as done
in \cite{Ade:2015hxq} but this complicates the relations to the
angular spectra, or work solely in the rotated complex variable $\bar
P$ which we leave for future work.} In what follows we calculate
$f_\ell^P$ using the same procedure described in \citet{Komatsu:2010fb}
but with $C_\ell^{TT}$ replaced by the total polarization power
$C_\ell^{EE}+C_\ell^{BB}$ when computing the moments of the
polarization field.

Under our assumption for $f_\ell^P$ the non-vanishing
combinations in equations (\ref{eq:ps1})-(\ref{eq:ps3}) simplify to
\begin{align}
  {\cal C}_+(\bm\theta) &=\int \frac{\ell
    d\ell}{2\pi} \,f_\ell^P \,\left(C_\ell^{E}
  +C_\ell^{B}\right)\,J_0(\ell\theta)\,,\label{eq:pp}\\
    {\cal C}_-(\bm\theta)&=\int \frac{\ell
    d\ell}{2\pi} \,f_\ell^P \,\left(C_\ell^{E}
  -C_\ell^{B}\right)\,J_4(\ell\theta)\,,\label{eq:pm}\\
  {\cal C}_\times(\bm\theta) &=2\,\int \frac{\ell
    d\ell}{2\pi}\,f_\ell^P \,C_\ell^{EB}\,J_4(\ell\theta)\,.\label{eq:px}
\end{align}
Two important points are readily apparent when considering
correlations of this form. Firstly, in equations
(\ref{eq:pp})-(\ref{eq:px}) we can see why the coordinate frame is not
a natural one for separating out $E$ and $B$-mode contributions. The
correlations ${\cal C}_+$ and ${\cal C}_-$ contain distinct linear
combinations of $C_\ell^{E}$ and $C_\ell^{B}$ but are filtered by
Bessel functions of different order. This means that we can only
separate $E$ and $B$-modes in the full-sky limit by using the
orthonormality of the basis
functions \citep{Kamionkowski:1996ks,Chon:2003gx} - the spin-2
spherical harmonics in the full-sky case.  In the flat-sky
approximation we cannot use the orthogonality properties of the Bessel
functions since the approximation breaks down at large angles. Thus
$E$ and $B$-modes are inevitably mixed when carrying out a partial-sky
analysis leading to well-known separation issues. If we are interested
in parity-violating modes however, ${\cal C}_\times$ provides a
perfectly good estimate of the amount of $EB$ correlation.

Secondly, the correlators are robust with respect to potential
systematics. For example, in equations (\ref{eq:ps1})-(\ref{eq:ps3})
we see that the contribution from mismatched peak biasing functions,
proportional to $f_\ell^-$, carry a specific dependence on the
azimuthal angle $\phi$ while the signal of interest, proportional to
$f^+_\ell$ in each case, is independent of $\phi$. The signal can
therefore be separated out by integrating over the $\phi$
dependence. In principle, this is the same for any contaminant that
is not described by statistically isotropic correlations of the form
shown in equations~(\ref{eq:cl1})-(\ref{eq:cl6}).

\begin{figure}[t]
\begin{center}\begin{tabular}{c}
    \includegraphics[width=8.5cm]{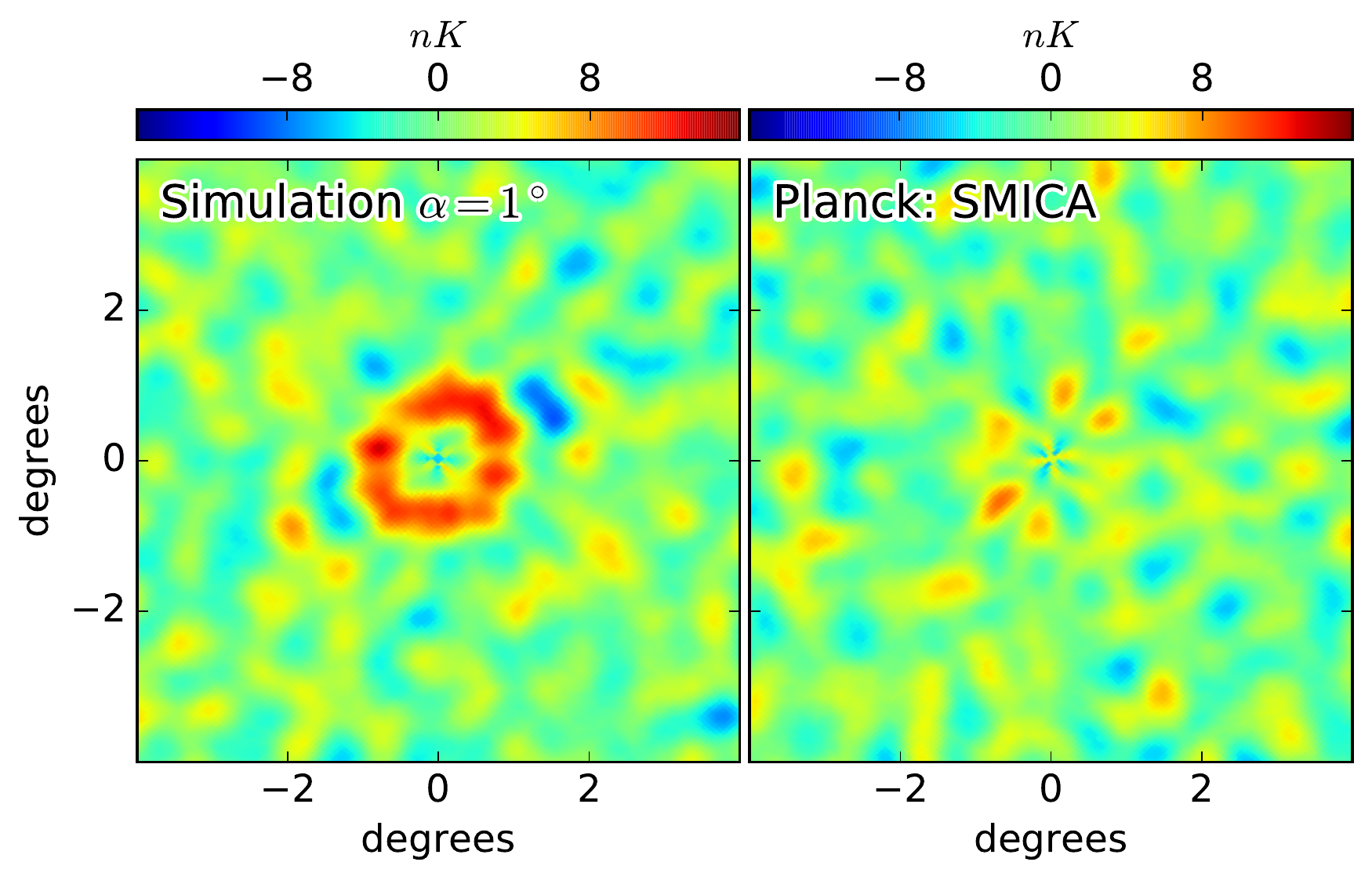}\\
    \includegraphics[width=8.5cm]{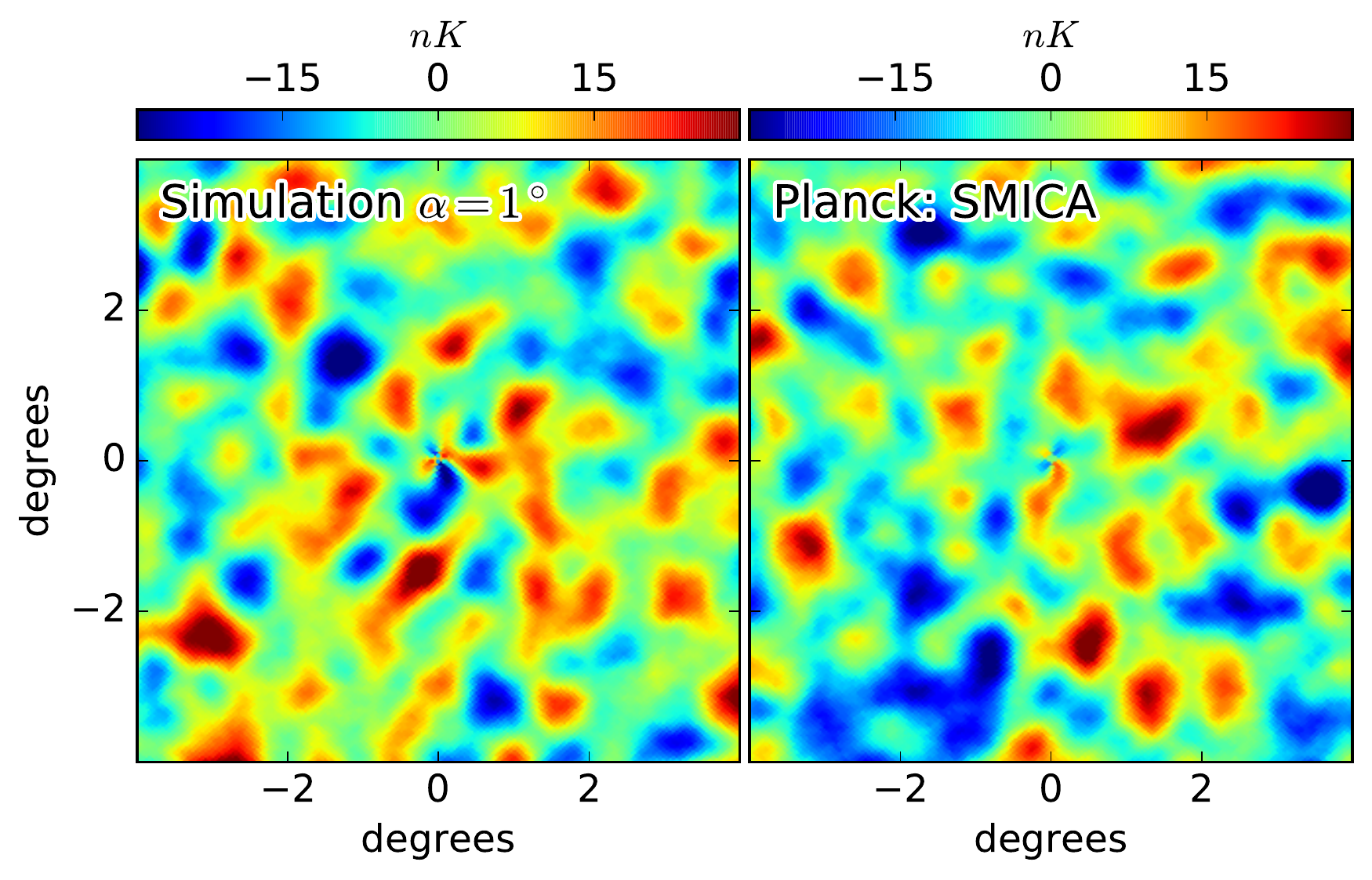}
\end{tabular}
\caption[]{Stacks for ${\cal C}_\times$ (top) and $\vev{U^T_r}$
  (bottom) for {\sl Planck} {\sl SMICA} maps compared to those from the
  $\alpha=1$ degree simulation. The {\sl SMICA} stack for ${\cal C}_\times$
  appears to show a weak radial signal consistent with a
  $J_4(\ell\theta)$ radial profile but with an additional azimuthal
  modulation. The background noise level in the $\vev{U^T_r}$ stack is too
  large to to identify a consistent $TB$ signal.}
  \label{fig:smica}
\end{center}
\end{figure}

\begin{figure*}
\begin{center}\begin{tabular}{cc}
\includegraphics[width=8cm]{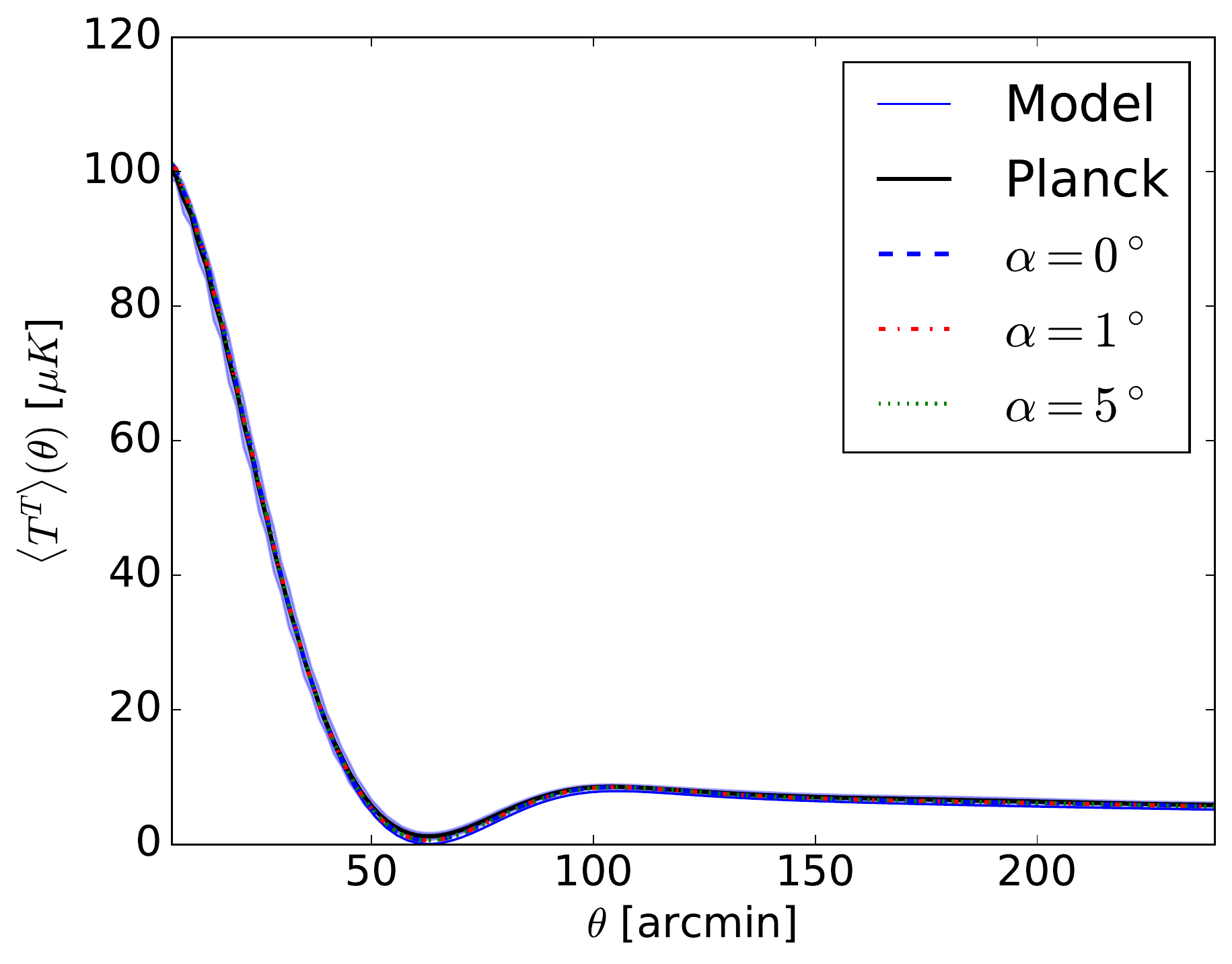}&
    \includegraphics[width=8cm]{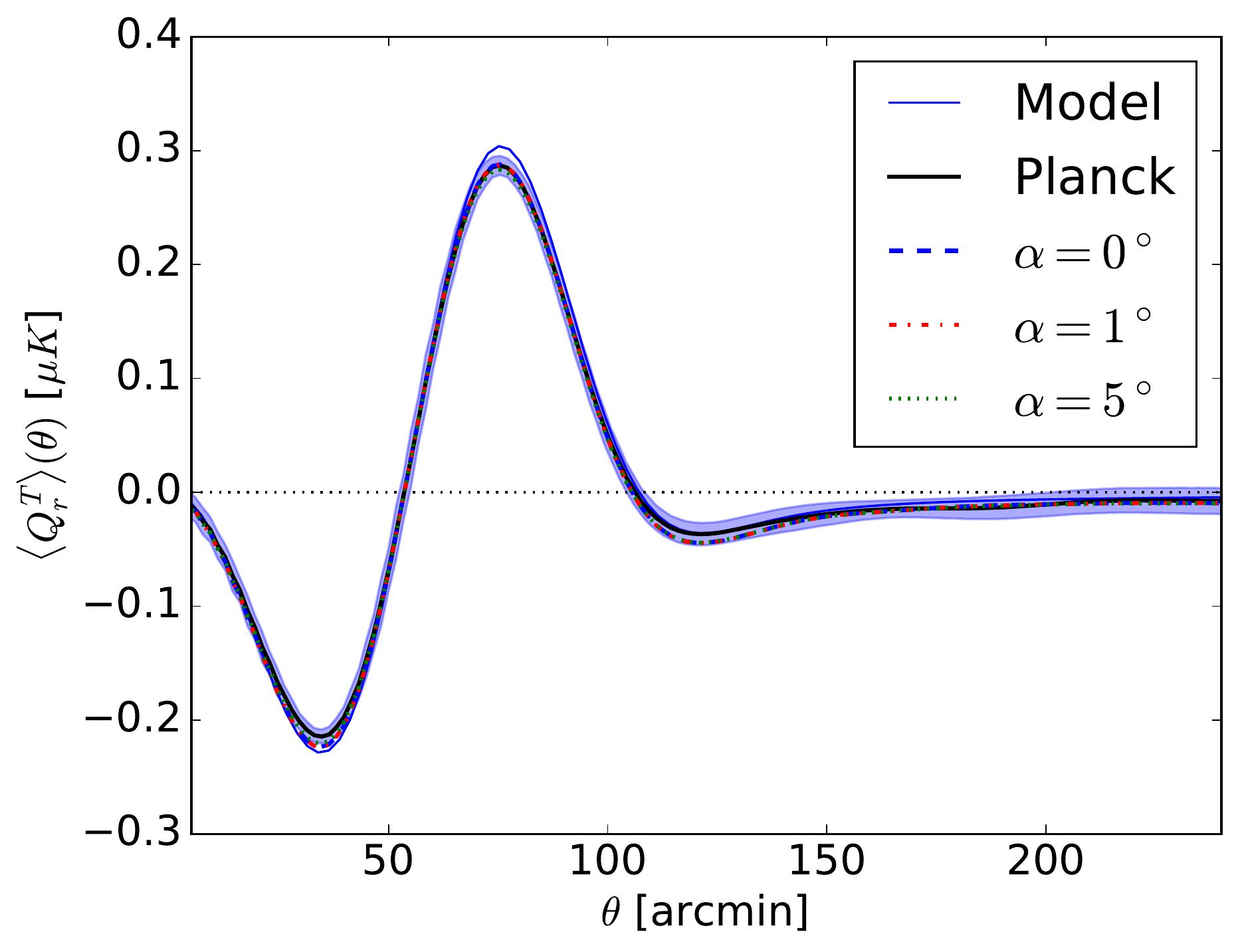} \\
    \includegraphics[width=8cm]{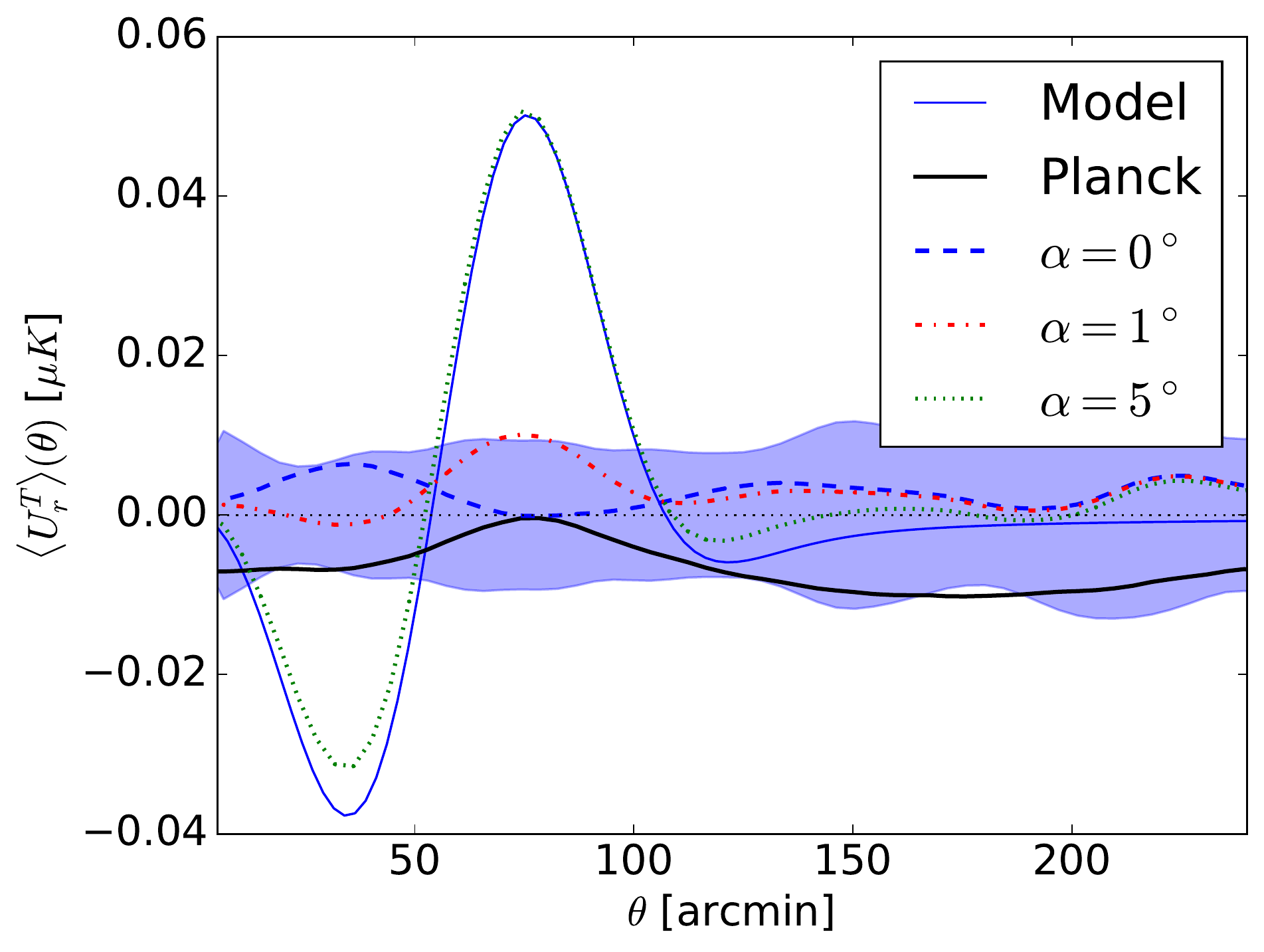}&
    \includegraphics[width=8cm]{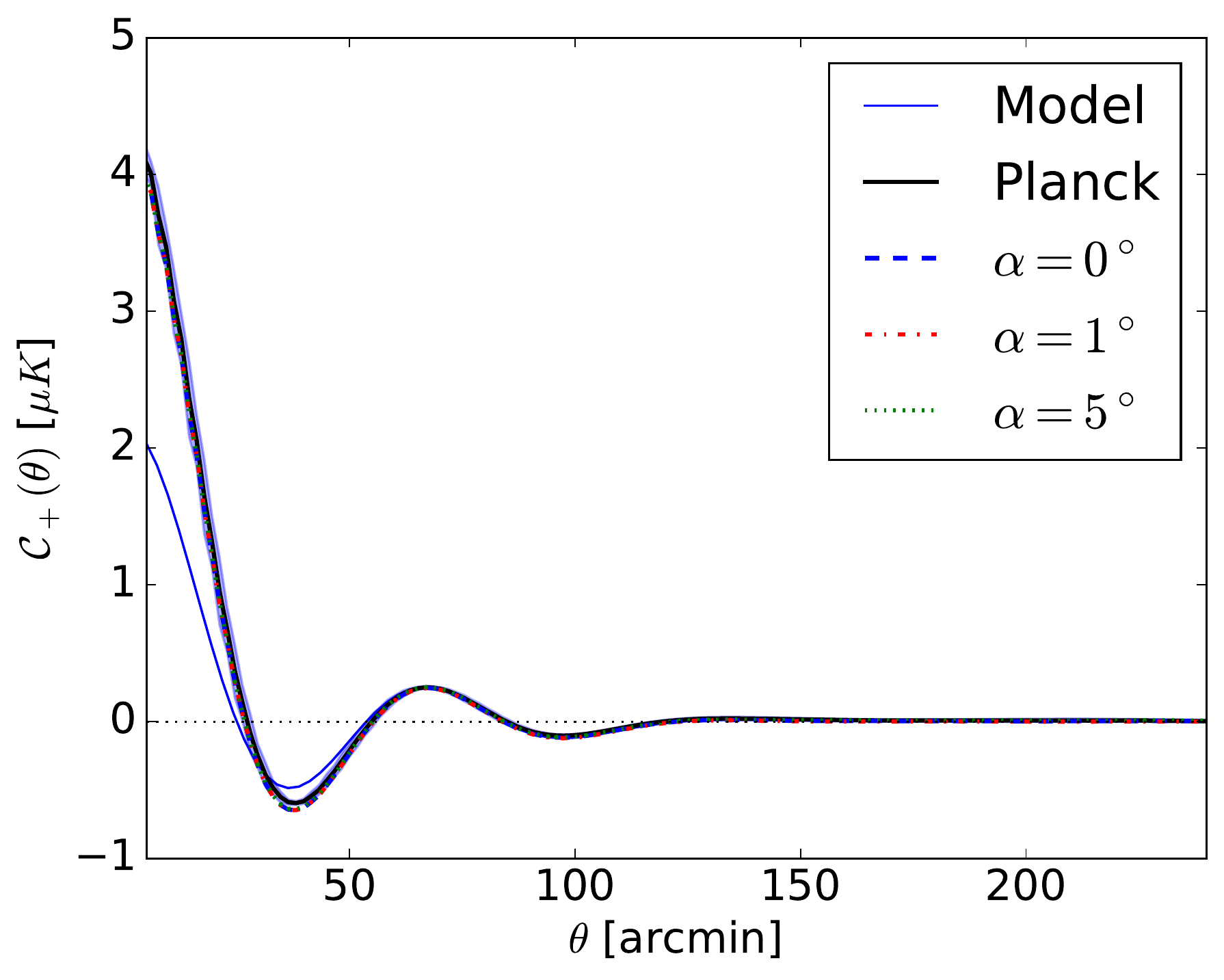} \\
    \includegraphics[width=8cm]{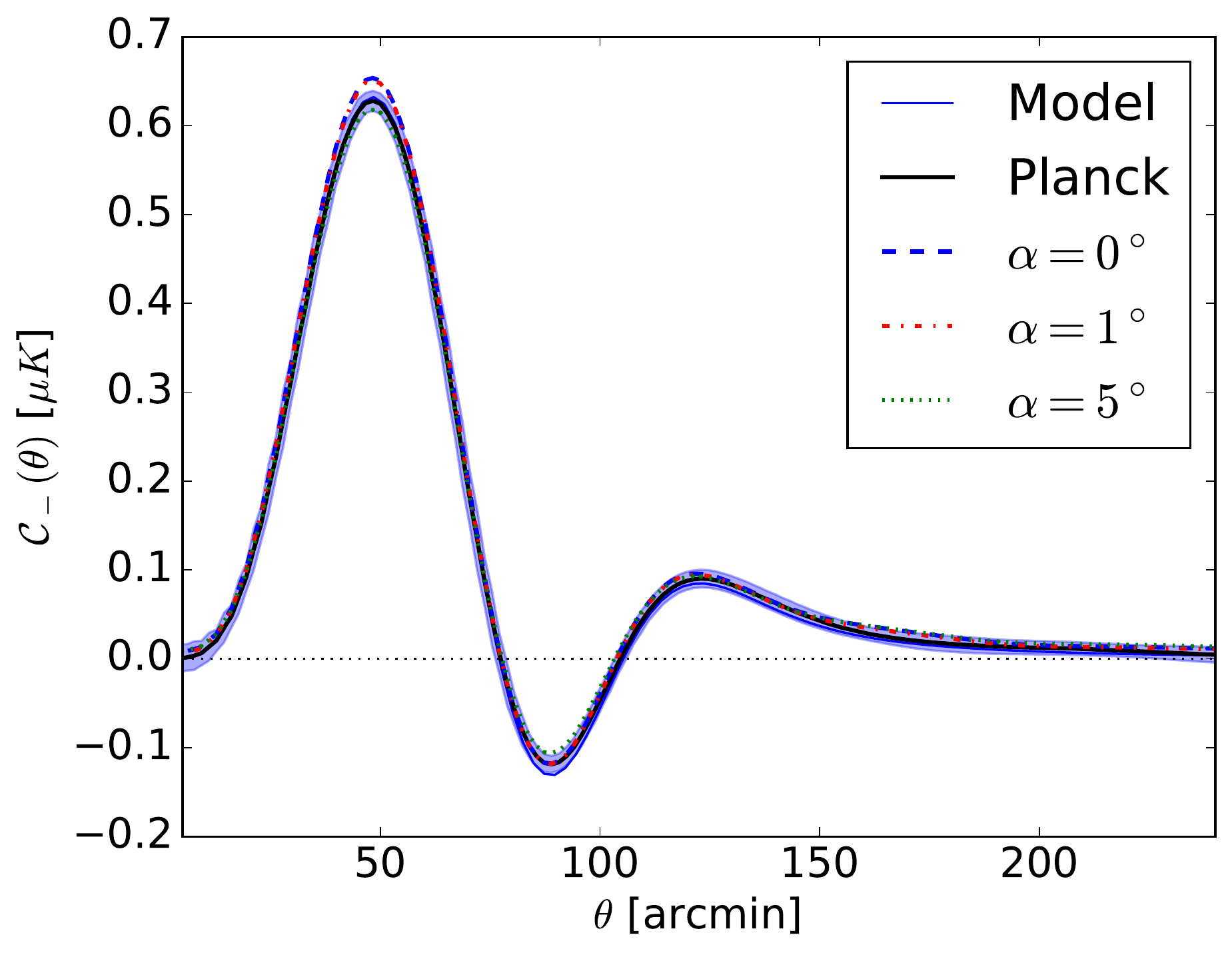}&
    \includegraphics[width=8cm]{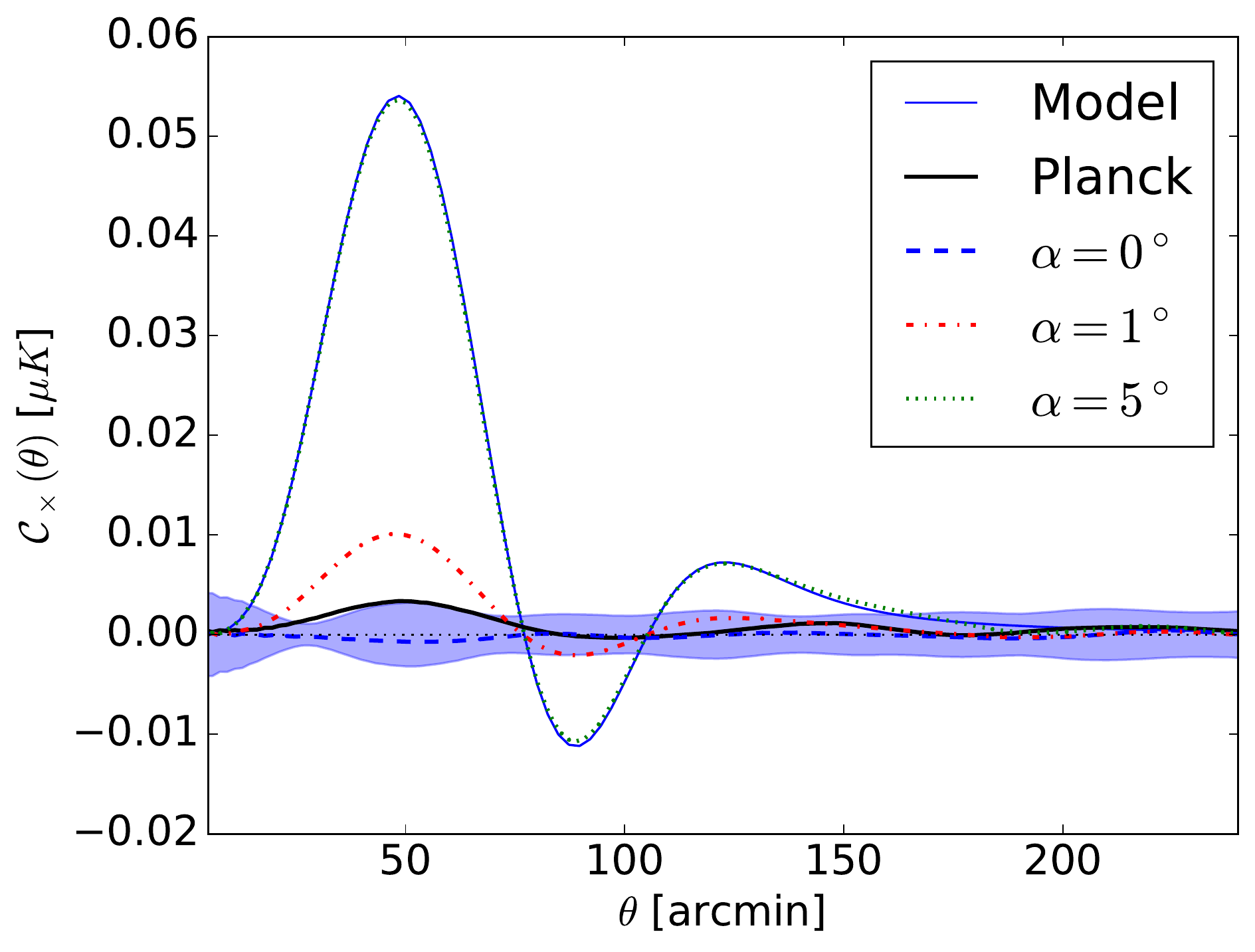}
\end{tabular}
\caption[]{Azimuthally averaged profiles of the stacks shown in
  figures~\ref{fig:simulations} and \ref{fig:smica}. For each panel we
  show the profiles for the three simulations and for the {\sl Planck}
  {\sl SMICA} map. The model curves are calculated using
  equations~(\ref{eq:qr})-(\ref{eq:ur}) and
  (\ref{eq:pp})-(\ref{eq:px}). We use a minimal $\Lambda$CDM \planck\
  best-fit, including tensors, as input power spectra and apply a
  rotation of $\alpha=5$ degrees. The model curves include only signal
  power. The shaded area shows the standard deviation of the profile
  calculated from an azimuthal averaging of the stack variances for
  the {\sl SMICA} maps centred on the {\sl SMICA} case except for the
  ${\cal C}_\times$ and $\vev{U^T_r}$ cases where no signal is
  expected. The model curves are a very accurate prediction of the
  observed signal. The ${\cal C}_+$, signal-only curve differs from
  the observed values at small angular scales where the noise bias of
  the polarization becomes significant.}
\label{fig:profiles}
\end{center}
\end{figure*}

\section{Application}\label{sec:appl}

We now implement two separate methods based on peak stacking as simple
examples of how we could use the correlators in
equations~(\ref{eq:qr})-(\ref{eq:ur}) and (\ref{eq:pp})-(\ref{eq:pm}) to constrain
parity-violating modes in the CMB. To test our methods we use
simulated CMB maps with parity-violating patterns induced by an
explicit rotation of the original polarization pattern by an overall
rotation angle $\alpha$. The rotation causes the mixing of power in
$T$, $E$, and $B$-modes and is described by the following angular
power spectra \citep{Lue:1998mq,Feng:2004mq}
\begin{align}
  &\tilde C_\ell^{T} = C_\ell^{T}\,,\label{eq:rot_spec}\\
  &\tilde C_\ell^{E} = C_\ell^{E} \cos^2(2\alpha) +
 C_\ell^{B}\sin^2(2\alpha)\,,\\
  &\tilde C_\ell^{B} = C_\ell^{E} \sin^2(2\alpha) +
  C_\ell^{B}\cos^2(2\alpha)\,,\\
  &\tilde C_\ell^{EB} =
  \frac{1}{2}\left(C_\ell^{E}-C_\ell^{B}\right)\sin(4\alpha)\,,\label{eq:rot_spec0}\\
  &\tilde C_\ell^{TE} = C_\ell^{TE}\cos(2\alpha)\,,\\
  &\tilde C_\ell^{TB} =C_\ell^{TE}\sin(2\alpha)\,,\label{eq:rot_spec1}
\end{align}
where  $C_\ell$ are the spectra of the original realization with
vanishing $EB$ and $TB$ correlations.

Our second method will be a constraint on a direction--dependent
rotation angle. To test the method we will use a simple toy model
where the rotation angle varies as a dipole across the sky. In both
cases we will also apply our estimation procedure to {\sl Planck} CMB
component maps.

\subsection{Average rotation}\label{sec:average}

We use maps pixelised using the {\tt
Healpix}\footnote{\url{http://healpix.sourceforge.net}}
\citep{Gorski:2004by} scheme at resolution $N_{\rm side}=1024$. We generate realizations of
the full-sky using the {\tt synfast} tool at this resolution with input
theoretical power spectra $C_\ell$ corresponding to the best-fit model
in ``base + r'' fits reported in the latest {\sl Planck} results
\citep{Ade:2015xua}. A second set of realizations is generated using the
rotated spectra $\tilde C_\ell$ in equations~(\ref{eq:rot_spec})-(\ref{eq:rot_spec1}). We generate two realizations
with rotation angles $\alpha=1$ and $\alpha=5$ degrees. All
realizations are smoothed using a $30$ arcminute Gaussian window
function and we have used the {\sl Planck} total intensity and
polarization transfer functions to smooth the theory to match the
experimental beam effects in all cases.

We also use the {\sl Planck} R2.0 {\sl SMICA} $I$, $Q$, and $U$ maps
\citep{Adam:2015tpy} down-graded to a common resolution of $N_{\rm side}=1024$ and smoothed by the
same $30$ arcminute Gaussian window function as the realizations. We
add the Half-Ring Half-Difference (HRHD) noise estimate to each of the
realizations although the noise is significantly suppressed at our
working resolution. Our polarization convention is different from that
used in the {\tt Healpix} package and corresponds to a change $U\to
-U$.

The analysis procedure is as follows: we run the {\tt Healpix} {\tt
hotspot} tool to identify all peaks (both minima {\sl and} maxima) in
each set of $I$, $Q$, and $U$ maps. We then take tiles, $8$ degrees
wide, centred at each peak location and rotate them so that the peak
is at a reference pixel. Each tile in $I$, $Q$, and $U$, is then added
to separate stack averages after the monopole of each tile is
subtracted \citep{Komatsu:2010fb,Ade:2015hxq}. The subtraction
minimises the effect of long-range correlations. We also exclude
regions inside the mask defined by the {\sl SMICA} confidence
parameter. The stacks we obtain are an average of some 36,000
locations including both minima and maxima in all three Stokes
parameters. We combine all maxima with thresholds above zero and the
{\sl negative} of all minima with thresholds below zero.

Figure~\ref{fig:simulations} shows the peak stacks in ${\cal C}_+$,
  ${\cal C}_-$, ${\cal C}_\times$, and $\vev{U^T_r}$ for simulations
with $\alpha=0$,  $1$, and $5$ degrees. For all cases the
${\cal C}_+$ and 
  ${\cal C}_-$ stacks, that are not sensitive to parity-violating
modes, display the expected radial envelope that decreases to the mean
field limit as we move away from the peak. ${\cal C}_\times$
and $\vev{U^T_r}$ only show a radial signal for the $\alpha\ne 0$
cases. For the $\alpha=1$ degree case the ${\cal C}_\times$ stack,
sensitive to $EB$ modes shows a clear signal whereas for the
respective 
$\vev{U^T_r}$ stack the signal is close to the background level
but clearly visible in the $\alpha=5$ degree case.

For both ${\cal C}_\times$
and $\vev{U^T_r}$ stacks, a signal with azimuthal dependence and
amplitude roughly comparable to the background noise is visible close
to the center of the stacks. This signal does not bias the estimate of
parity-violation since it averages to zero when integrating in azimuthal
angle $\phi$. 

The azimuthally averaged profiles for the stacks are shown in
Figure~\ref{fig:profiles}. The stacks from simulated maps are
consistent with the expected Bessel functions of zeroth and fourth
order while the amplitude grows as a function of overall rotation
$\alpha$. The figure shows an estimate of the background noise
calculated using the standard deviation obtained by binning the
variance of the stacks using the same procedure as the stacks
themselves. In all cases the stacks that are only sensitive to the
sum and difference $EE$ and $BB$ modes show highly
significant signals. The signal from $EB$ modes is significantly above
the noise level for both $\alpha\ne 0$ simulations in the ${\cal
C}_\times$ stack.  The $TB$ modes do not show up as strongly compared
to the noise which is larger due to the large background induced by
the correlation with the total intensity.

Figure~\ref{fig:profiles} also includes model predictions based on the
minimal $\Lambda$CDM {\sl Planck} best-fit model including
tensors. The models are in excellent agreement with all the curves
except for the ${\cal
 C}_+$ case which peaks at $\theta=0$ and is sensitive to the residual
 noise bias. We have checked that adding an arbitrary white noise
 component to the model recovers the observed curve in this case. The
 agreement between the models and the observed and simulated curves
 validates our analytic treatment of the peak constrained stacks and
 in particular shows that our assumptions for the calculation of the
 $f^P_\ell$ bias function are robust. 

The {\sl Planck} {\sl SMICA} map shows a hint of $EB$ signal at a
 level $< 1$ degree. It is also consistent with a $J_4(x)$ profile for
 $\theta<100$ arcminutes which is an indication that it may be sourced by an
 overall rotation. The $\vev{U^T_r}$ mode is noisier than the ${\cal
 C}_\times$ channel. We can attempt to obtain an estimate for the
 overall rotation angle $\alpha$ by fitting rotated models
 (\ref{eq:rot_spec0}) and (\ref{eq:rot_spec1}) to the {\sc Planck}
 azimuthally averaged curves. As an estimate of the error in ${\cal
 C}_\times$ we use the variance maps accumulated during the stacking
 and co-added using the same azimuthal averaging. The correlations in
 the angular bins shown in Figure~\ref{fig:profiles} are large. This
 can also be seen in the correlations, at the smoothing scale $\sim
 30$ arminutes, of the noise fluctuations far from
 the central peaks in the stacked images. We therefore obtain our
 estimates using a simple $\chi^2$ fit assuming uncorrelated errors
 but on a much coarser binning of the azimuthally averaged profiles to
 minimize the correlations. We find that the estimate is stable for
 bins larger than $10$ arcminutes.

For a bin size equal to half the smoothing
 scale we obtain an joint estimate from fits to $\vev{U^T_r}$ and ${\cal
 C}_\times$ of $\alpha < 0.72$ degrees at $95\%$ confidence. This is in broad
agreement with a recent analysis using {\sl Planck} data
by \cite{Gruppuso:2015xza}.

\subsection{Direction-dependent rotation}

\begin{figure}
\begin{center}\begin{tabular}{c}
    \includegraphics[width=8cm]{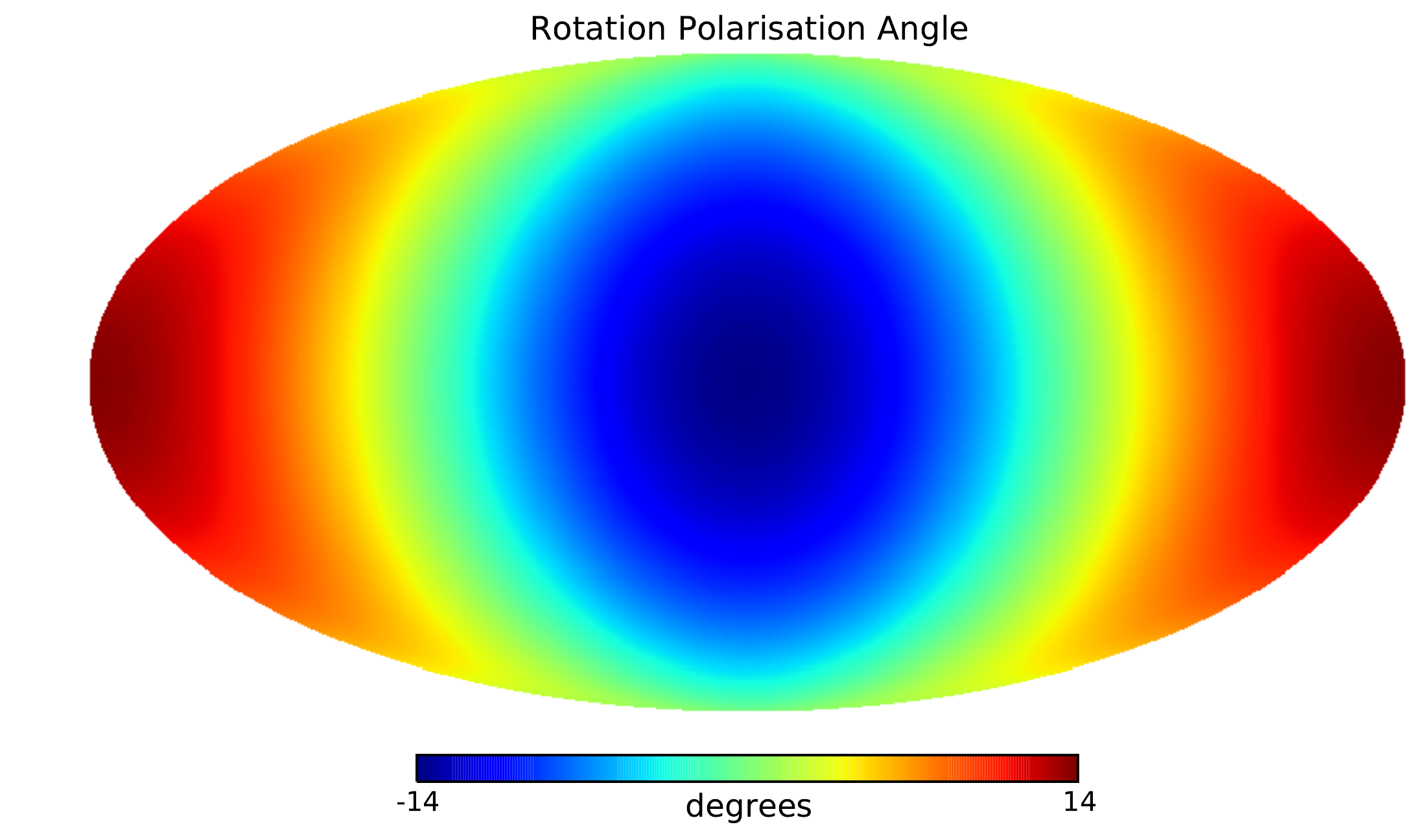}\\
    \includegraphics[width=8cm]{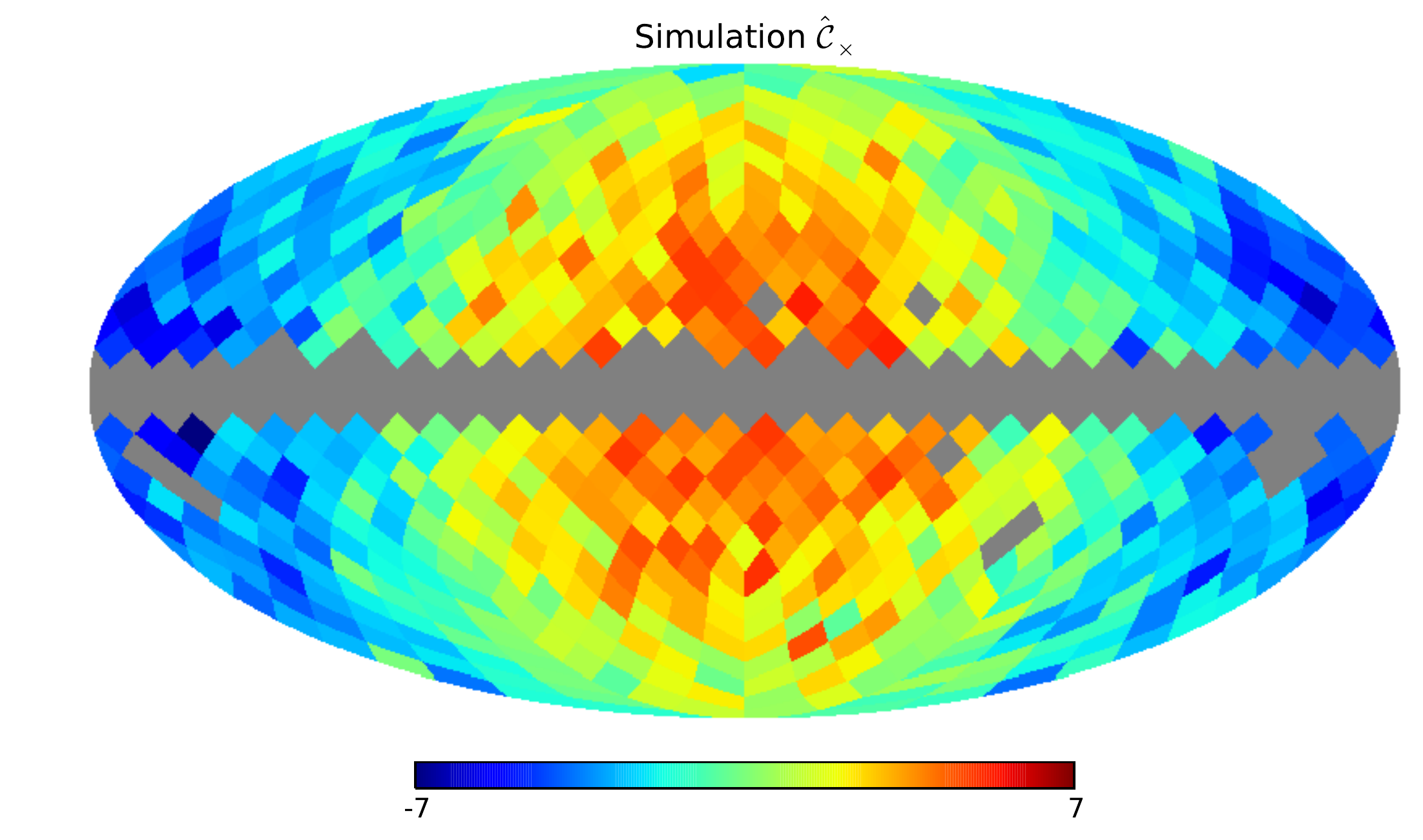} \\
    \includegraphics[width=8cm]{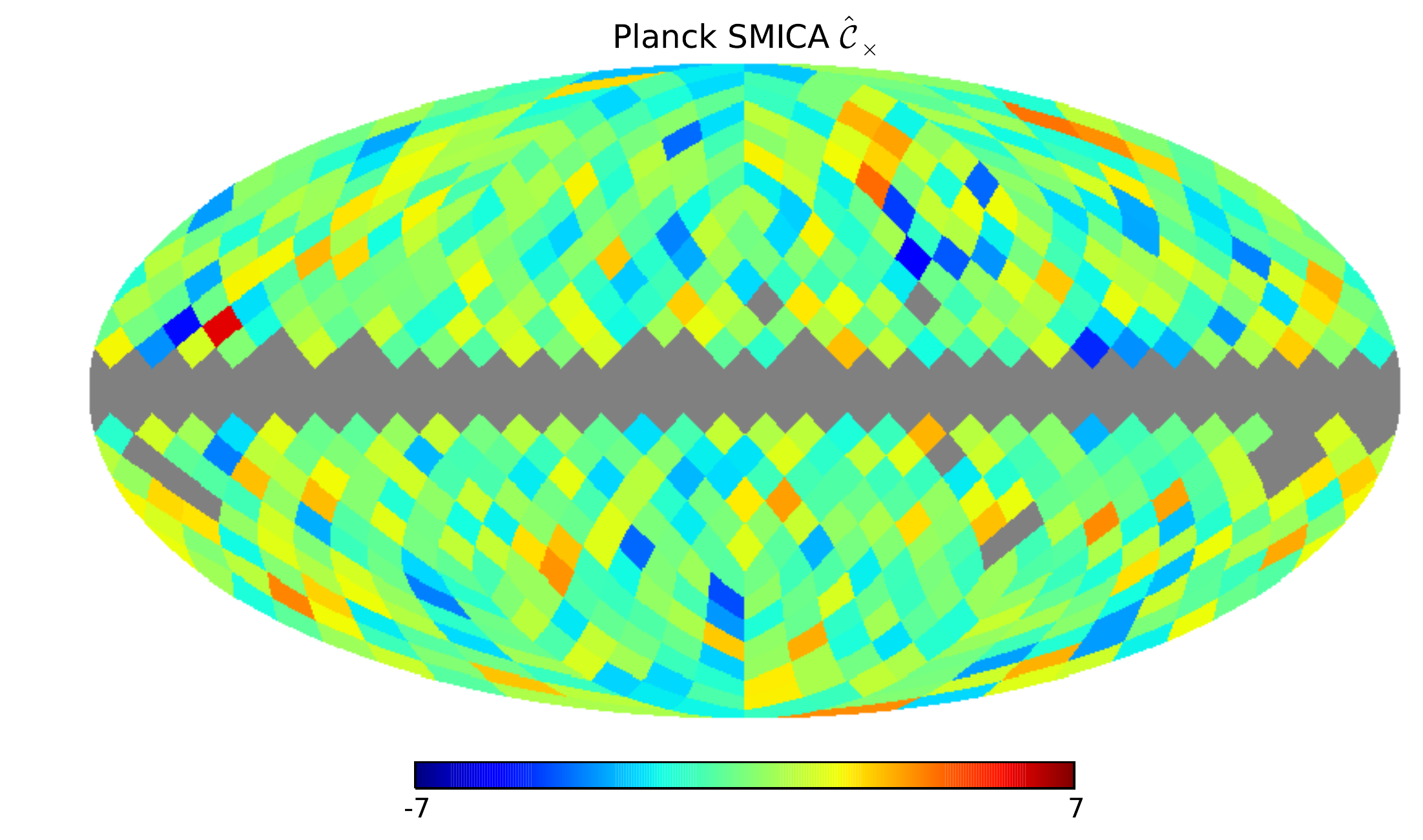}
\end{tabular}
  \caption[]{Map of input rotation angle  $\alpha(\hat {\bm n})$ toy
    model ({\sl
    top}) and the result of running our binned 
    $\hat{{\cal C}}_\times$ estimation pipeline for a {\sl Planck}-like simulation with
    polarization rotated by the input map at each pixel ({\sl
    middle}), {\sl Planck} {\sl SMICA}
    map ({\sl bottom}). 
  }
\label{fig:alpha_maps}
\end{center}
\end{figure}

An analysis using peak stacks is not restricted to constraining the
monopole of a possible rotation. In fact, if birefringence exists, there
is no reason why it should induce a coherent rotation of the
polarization angle over the entire sky. Recent constraints on
birefringent effects have employed multipole based estimators for the
direction dependent angle
\citep{Ade:2015cao,Gluscevic:2012me} based on the estimator developed by
\citep{Gluscevic:2009mm}. The methods result in an estimate of the
monopole expansion of the angle $\alpha_\ell$. These estimates based on measures of rotation in
multipole space face the same problems with respect to inhomogeneous
weighting and partial-sky coverage as conventional power spectrum
estimation methods. In contrast an estimate based on peak stacks can
provide a map of any potential signal without complications due to sky
coverage and is only limited by the fact that peak stacking is not an
optimal compression of {\sl all} the information in the maps. It also
has the advantage, shared with any coordinate space based measure,
that it is more suitable to detect any signal that is {\sl compact} on the
sky. This makes it complementary to multipole based methods that are
ideally suited to detect signals that are centred around certain
{\sl angular} scales or that have well defined, statistically
isotropic, signatures. 

As an example of how one could employ coordinate space based
correlation function methods to constrain $\alpha(\hat {\bm n})$ we
modify our peak stacking pipeline to consider averages inside lower
resolution pixels.The toy model we use to test this method is obtained
by generating a map $\alpha(\hat {\bm n})$ using the {\tt Healpix} {\tt alm2map}
tool with a pure ($\ell=1$, $m=1$) dipole mode with
$\alpha_{11}=20$ degrees as input. We use the map to apply a pixel
dependent rotation of the polarization in the simulated {\sl Planck}
map with no initial parity-violating modes. The peak stacking
procedure described in section~\ref{sec:average} is repeated but the
results are binned by low resolution pixels of $N_{\rm side}=8$. We
then use the peak value of the ${\cal C}_\times$ estimate in each
pixel, ${\hat{\cal C}}_\times = {\rm max}(|{\cal C}_\times(\theta)|)$, as a
tracer of the amount of average rotation in each low resolution pixel
since ${\cal C}_\times$ gives the highest signal-to-noise
determination of the rotation. 

\begin{figure}
\begin{center}
    \includegraphics[width=9cm]{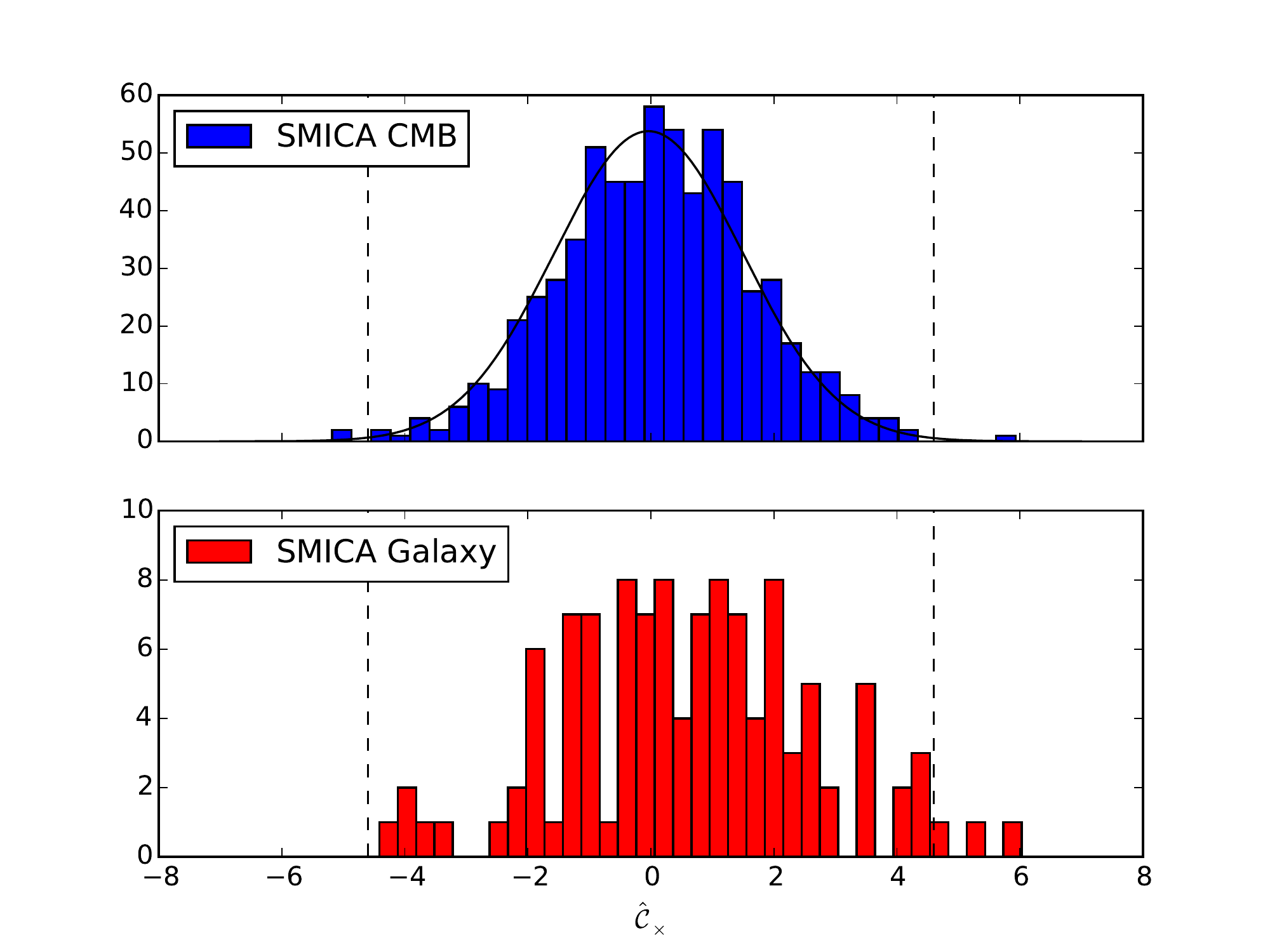}
  \caption[]{Histograms of ${\hat{\cal C}}_\times$ values found in the
  {\sl SMICA} map outside ({\sl top}) and inside ({\sl bottom}) the
  confidence (galactic) exclusion mask. The vertical lines indicate
  the $\pm 3\sigma$ bounds of the Gaussian which gives the best-fit to
  the distribution of pixels outside the mask. All pixels outside the
  $3\sigma$ bound are either inside the mask or close to the boundary
  pointing to a correlation with foregrounds.
  }
\label{fig:alpha_histo}
\end{center}
\end{figure}

The results of this exercise are summarised in
Figure~\ref{fig:alpha_maps} which shows how ${\hat{\cal C}}_\times$
traces the input $\alpha(\hat {\bm n})$ in the simulation. We also
show the result of running this estimate on the  {\sl SMICA}
map itself. The $\alpha(\hat {\bm n})$ map for {\sl SMICA} does not
show any obvious signal although it is interesting to note that the
two pixels at the extremes of the range in  ${\hat{\cal C}}_\times$
lie close to the excluded region that fails the {\sl SMICA} confidence
criterion. We leave a full analysis of the significance of these
estimates for future work but in Figure~\ref{fig:alpha_histo} we show
the binned distribution of pixel values outside and inside of the
excluded region. All pixels that lie outside a $\pm 3\sigma$ bound
defined by a best-fit Gaussian to the excluded pixels lie either
inside the masked area or are immediately adjacent to it. The variance
in the map also appears to be anisotropic with minima around the
equatorial poles which correlates with the variance due to {\sl
Planck}'s scan strategy.

\section{Discussion}\label{sec:disc}

We have shown how correlations of Stokes parameters of CMB maps can
yield robust, local estimates of parity-violating signals. This can be
useful when analysing maps to search for localised signals for which
complementary harmonic based estimators are not ideal. Techniques
based on coordinate frame correlation function estimators are
therefore similar in spirit to many techniques used to investigate
possible localised anomalies in the CMB (see \cite{Ade:2015xua} for a
recent overview) which rely on coordinate frame statistics.

We used the correlators defined here to build simple tools to image
parity-violating effects. We focused on peak stacking techniques as
these provide a powerful compression of the acoustic signature in the
polarization {\sl and} are relatively easy to compute as opposed to
the full, unconstrained correlations at the same resolution
level. We have found the signal of parity-violation appears as expected in
simulations where a non-zero effect was included and that our analytic
estimates of the signal are in good agreement with the simulations. Our estimators
run on {\sl Planck} maps yield a limit on a sky-averaged rotation angle that is
compatible with estimates in the literature obtained using different methods.

Our methods are not limited to the small angle limit we have adopted
in this work. Peak stacking techniques focus are optimal for signals
that modify the correlations on acoustic scales around a degree on the
sky. They are well suited for along the line-of-sight birefringence
effects. Other classes of parity-violating models however, may induce
signals on larger angular scales if the effects were active during the
inflationary phase in the early universe. In future work we will
extend our methods to the full-sky. An additional advantage of this
will be the exploitation of the full information in the maps. This
will increase the signal-to-noise of any estimate and allows for much
higher resolution reconstructions of maps such as those shown in
Figure~\ref{fig:alpha_maps}.

We used a simple exercise to show how the localised nature of these
methods can be harnessed to reconstruct maps of spatially varying
rotation angle. The example used a very large input rotation dipole as
the signal to be reconstructed and the resulting map is much lower
resolution than the input {\sl Planck} maps. The choice of resolution
was dictated by the limited number of peaks in the maps and we expect
similar applications using the full-sky information will yield much
higher resolution reconstructions.

Another advantage of methods based on coordinate frame correlations is
the relatively straightforward implementation of noise weighting. We
have not explored this here but we expect the results would be improved
by correctly noise weighting the correlation estimators since the
polarization information in current maps is still in the relatively
low signal-to-noise regime. This would also help in targeting smaller
angular resolutions in current and near-term polarization data.

On the small scales an interesting prospect is to explore how
localised, peak stack estimators based on correlation functions of the
polarization could help to extract {\sl lensing} information from CMB
maps. Gravitational lensing of CMB photons, now routinely measured
(see \cite{Ade:2015zua} and references therein for a summary), leaves
an imprint in the polarization signal that mixes $E$ and $B$
modes. Current estimates are based on higher-order correlations in
harmonic space \citep{Okamoto:2003zw} and real
space \citep{Carvalho:2010rz,Bucher:2010iv}. Complementary techniques similar to
those developed in this work may be feasible. In particular peak
stacking may be ideally suited to isolate the lensing signal in a
manner that is reminiscent of exploratory methods used to isolate the
CMB lensing signal due to
halos \citep{Baxter:2014frs,Madhavacheril:2014slf}.

\acknowledgments

This work is supported by the Science and Technology Facilities Grant ST/J0003533/1.

\bibliography{refs}
\end{document}